\documentclass[aps,a4paper,10pt,twocolumn,footinbib]{revtex4} 
\usepackage[T1]{fontenc}
\usepackage{mathptmx}
\usepackage{datetime}
\usepackage{amsmath}
\usepackage{amsfonts}
\usepackage{amsfonts}
\usepackage{mathrsfs}
\usepackage[mathscr]{euscript}
\usepackage[dvips]{graphicx}
\usepackage{fancyhdr}
\usepackage{colordvi}
\usepackage{hyperref} 
\usepackage{epsfig}
\usepackage{color}
\usepackage{bm}
\usepackage{tikz}

\def\FIGWIDTH{0.840}

\usepackage{tikz}
\usepackage{color}

\def\innerprod(#1,#2){{\left<#1\,,\,#2\right>}}

\newcommand{\figlabelsize}[1]{\huge{#1}}

\def\mm{{\textup{mm}}}

\begin{document}
\title{Mode profile shaping in wire media: 
Towards an experimental verification}

\author{Taylor Boyd$^{1,2,3}$}

\author{Jonathan Gratus$^{2,1}$}
\homepage[]{https://orcid.org/0000-0003-1597-6084}
\email[\hphantom{.}~]{j.gratus@lancaster.ac.uk}

\author{Paul Kinsler$^{2,1}$}
\homepage[]{https://orcid.org/0000-0001-5744-8146}
\email[\hphantom{.}~]{Dr.Paul.Kinsler@physics.org}

\author{Rosa Letizia$^{3,1}$}
\homepage[]{https://orcid.org/0000-0002-1664-2265}
\email[\hphantom{.}~]{r.letizia@lancaster.ac.uk}

\author{Rebecca Seviour$^{4}$}
\homepage[]{https://orcid.org/0000-0001-8728-1463}
\email[\hphantom{.}~]{r.seviour@hud.ac.uk}

\affiliation{
    $^1$Cockcroft Institute, 
    Sci-Tech Daresbury, 
    Daresbury WA4 4AD, 
    United Kingdom.
}

\affiliation{
  $^2$Physics Department,
  Lancaster University,
  Lancaster LA1 4YB, 
  United Kingdom.
}

\affiliation{
  $^3$Engineering Department,
  Lancaster University,
  Lancaster LA1 4YB, 
  United Kingdom.
}

\affiliation{
  $^4$University of Huddersfield,
  Huddersfield HD1 1JB,
  United Kingdom.
}

\chead{Mode profile shaping in wire media}
\rhead{
}

\begin{abstract}
We show that an experimentally plausible system 
 consisting of a modulated wire medium hosted in a metal cavity 
 can preserve the longitudinal field profile shaping 
 predicted by Boyd et al. (2018)
 on the basis of a perfectly periodic wire-only structure.
These new frequency domain numerical results are a significant step
 towards justifying the construction of 
 an experimental apparatus to test the field profile shaping
 in practise.
\end{abstract}


\maketitle

\section{Introduction}
\label{ch_Intro}

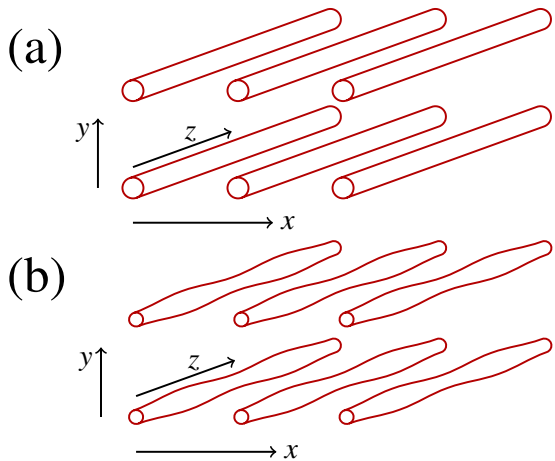
\begin{figure}[b]
\centering
\resizebox{\FIGWIDTH\columnwidth}{!}{
\raisebox{2.5cm}{\figlabelsize{(a)}}
\begin{tikzpicture}
\draw [red!70!black,shift={(0,0)},thick] (0,0) circle (.15) (110:0.15) -- +(20:3)
    (-70:0.15) -- +(20:3) arc (-70:110:0.15) ;
\draw [red!70!black,shift={(1.5,0)},thick] (0,0) circle (.15) (110:0.15) -- +(20:3)
    (-70:0.15) -- +(20:3) arc (-70:110:0.15) ;
\draw [red!70!black,shift={(3,0)},thick] (0,0) circle (.15) (110:0.15) -- +(20:3)
    (-70:0.15) -- +(20:3) arc (-70:110:0.15) ;
\draw [red!70!black,shift={(0 ,1.4)},thick](0,0) circle (.15) (110:0.15) -- +(20:3)
    (-70:0.15) -- +(20:3) arc (-70:110:0.15) ;
\draw [red!70!black,shift={(1.5 ,1.4)},thick](0,0) circle (.15) (110:0.15) -- +(20:3)
    (-70:0.15) -- +(20:3) arc (-70:110:0.15) ;
\draw [red!70!black,shift={(3 ,1.4)},thick](0,0) circle (.15) (110:0.15) -- +(20:3)
    (-70:0.15) -- +(20:3) arc (-70:110:0.15) ;
\draw [->,thick] (-0.5,0) +(0,.8) node[left] {\large $y$} +(0,0) -- +(0,1) ;
\draw [->,thick] (0,-0.5) +(2.0,0) node[right] {\large $x$} +(0,0) -- +(2,0) ;
\draw [->,thick] (0,0.3) +(30:.95) node {\large $z$} +(0,0) -- +(20:1.5) ;

\end{tikzpicture}}
\\
\resizebox{\FIGWIDTH\columnwidth}{!}{
\raisebox{2.5cm}{\figlabelsize{(b)}}
\begin{tikzpicture}
\draw [red!70!black,shift={(0,0)},thick,rotate=20] (0,0) circle (.10)
    (0,0.10) to[out=0,in=180] (.75,.16) to[out=0,in=180] (1.5,.1)
    to[out=0,in=180] (2.25,.16) to[out=0,in=180] (3,.1)
    (0,-0.10) to[out=0,in=180] (.75,-.16) to[out=0,in=180] (1.5,-.1)
    to[out=0,in=180] (2.25,-.16) to[out=0,in=180] (3,-.1) arc (-90:90:.1) ;
\draw [red!70!black,shift={(1.5,0)},thick,rotate=20] (0,0) circle (.10)
    (0,0.10) to[out=0,in=180] (.75,.16) to[out=0,in=180] (1.5,.1)
    to[out=0,in=180] (2.25,.16) to[out=0,in=180] (3,.1)
    (0,-0.10) to[out=0,in=180] (.75,-.16) to[out=0,in=180] (1.5,-.1)
    to[out=0,in=180] (2.25,-.16) to[out=0,in=180] (3,-.1) arc (-90:90:.1) ;
\draw [red!70!black,shift={(3,0)},thick,rotate=20] (0,0) circle (.10)
    (0,0.10) to[out=0,in=180] (.75,.16) to[out=0,in=180] (1.5,.1)
    to[out=0,in=180] (2.25,.16) to[out=0,in=180] (3,.1)
    (0,-0.10) to[out=0,in=180] (.75,-.16) to[out=0,in=180] (1.5,-.1)
    to[out=0,in=180] (2.25,-.16) to[out=0,in=180] (3,-.1) arc (-90:90:.1) ;
\draw [red!70!black,shift={(0,1.4)},thick,rotate=20] (0,0) circle (.10)
    (0,0.10) to[out=0,in=180] (.75,.16) to[out=0,in=180] (1.5,.1)
    to[out=0,in=180] (2.25,.16) to[out=0,in=180] (3,.1)
    (0,-0.10) to[out=0,in=180] (.75,-.16) to[out=0,in=180] (1.5,-.1)
    to[out=0,in=180] (2.25,-.16) to[out=0,in=180] (3,-.1) arc (-90:90:.1) ;
\draw [red!70!black,shift={(1.5,1.4)},thick,rotate=20] (0,0) circle (.10)
    (0,0.10) to[out=0,in=180] (.75,.16) to[out=0,in=180] (1.5,.1)
    to[out=0,in=180] (2.25,.16) to[out=0,in=180] (3,.1)
    (0,-0.10) to[out=0,in=180] (.75,-.16) to[out=0,in=180] (1.5,-.1)
    to[out=0,in=180] (2.25,-.16) to[out=0,in=180] (3,-.1) arc (-90:90:.1) ;
\draw [red!70!black,shift={(3,1.4)},thick,rotate=20] (0,0) circle (.10)
    (0,0.10) to[out=0,in=180] (.75,.16) to[out=0,in=180] (1.5,.1)
    to[out=0,in=180] (2.25,.16) to[out=0,in=180] (3,.1)
    (0,-0.10) to[out=0,in=180] (.75,-.16) to[out=0,in=180] (1.5,-.1)
    to[out=0,in=180] (2.25,-.16) to[out=0,in=180] (3,-.1) arc (-90:90:.1) ;
\draw [->,thick] (-0.5,0) +(0,.8) node[left] {\large $y$} +(0,0) -- +(0,1) ;
\draw [->,thick] (0,-0.5) +(2.0,0) node[right] {\large $x$} +(0,0) -- +(2,0) ;
\draw [->,thick] (0,0.3) +(30:.95) node {\large $z$} +(0,0) -- +(20:1.5) ;
\end{tikzpicture}}
\caption{
A wire medium is formed from an array of parallel metal or dielectric wires. 
In this work we use results based on rectangular arrays of
 (a) wires with uniform radii
 to predict the parameters needed for customised
 (b) wires with varying radii
 that generate a desired subwavelength field profile shaping.}
\label{fig_wires}
\end{figure}

Spatial dispersion is a valuable tool,
 which can be used to customise the profiles
 of electric fields in metamaterial and photonic structures.  
As we have already shown
 \cite{Gratus-KLB-2017apa-malaga,Gratus-KLB-2017jpc,Boyd-GKL-2018oe-tbwire},
 such field profile customisation is even possible
 at the sub-wavelength scale, 
 and does not require exhaustive brute-force computation.  
Further,
 for our spatially dispersive wire medium \cite{Boyd-GKL-2018oe-tbwire}, 
 we show this system can support
 shaped \emph{longitudinal} electric fields.  
This field profiling is achieved by varying the radius of the wires
 in a carefully calibrated way. 
In \cite{Boyd-GKL-2018oe-tbwire} the system was numerically modelled
 as an infinite periodic array of wires,
 so that the computation only required a single wire
 with periodic boundary conditions.  
In addition to periodicity,
 the simulated wire material had a very high permittivity
 and concomitantly small radius, 
 features that would be problematic in experiment, 
 both in terms of manufacture and fragility.  
Nevertheless, 
 we did show that such extreme structures were not necessarily required, 
 since re-scaled calculations indicated the shaping effect
 would still persist even with lower permittivities and larger wire radii.

\begin{figure}[t]
\centering
\resizebox{\FIGWIDTH\columnwidth}{!}{
\begin{tikzpicture}

\filldraw [fill=black!60!white,draw=black,fill opacity=0.2,very thick,line join=bevel,shift={(20:2)}]
   (-0.3,-0.2) rectangle (4.8,3.2) ;
\filldraw [fill=black!30!blue,fill opacity=0.2,draw=black,very thick,line join=bevel]
   (-0.3,-0.2) -- ++(20:2) -- ++(90:3.4)  -- ++(20:-2) -- cycle ;
\filldraw [fill=black!30!blue,fill opacity=0.2,draw=black,very thick,line join=bevel]
   (-0.3,-0.2) -- ++(20:2) -- ++(0:5.1)  -- ++(20:-2) -- cycle ;

\draw [draw=black,fill=red!90!black,shift={(0,0)},thick,rotate=20]
    (0,0) circle (.10)
    (0,0.10) to[out=0,in=180]  (1,.17)  to[out=0,in=180] (2,.1)
    (0,0.10) arc (90:-70:.1) to[out=0,in=180]  (1,-.17)
    to[out=0,in=180] (2,-.1) arc (-90:90:.1) ;
\draw [draw=black,fill=red!90!black,shift={(1.5,0)},thick,rotate=20]
    (0,0) circle (.10)
    (0,0.10) to[out=0,in=180]  (1,.17)  to[out=0,in=180] (2,.1)
    (0,0.10) arc (90:-70:.1) to[out=0,in=180]  (1,-.17)
    to[out=0,in=180] (2,-.1) arc (-90:90:.1) ;
\draw [draw=black,fill=red!90!black,shift={(3,0)},thick,rotate=20]
    (0,0) circle (.10)
    (0,0.10) to[out=0,in=180]  (1,.17)  to[out=0,in=180] (2,.1)
    (0,0.10) arc (90:-70:.1) to[out=0,in=180]  (1,-.17)
    to[out=0,in=180] (2,-.1) arc (-90:90:.1) ;
\draw [draw=black,fill=red!90!black,shift={(4.5,0)},thick,rotate=20]
    (0,0) circle (.10)
    (0,0.10) to[out=0,in=180]  (1,.17)  to[out=0,in=180] (2,.1)
    (0,0.10) arc (90:-70:.1) to[out=0,in=180]  (1,-.17)
    to[out=0,in=180] (2,-.1) arc (-90:90:.1) ;

\draw [draw=black,fill=red!90!black,shift={(0,1)},thick,rotate=20]
    (0,0) circle (.10)
    (0,0.10) to[out=0,in=180]  (1,.17)  to[out=0,in=180] (2,.1)
    (0,0.10) arc (90:-70:.1) to[out=0,in=180]  (1,-.17)
    to[out=0,in=180] (2,-.1) arc (-90:90:.1) ;
\draw [draw=black,fill=red!90!black,shift={(1.5,1)},thick,rotate=20]
    (0,0) circle (.10)
    (0,0.10) to[out=0,in=180]  (1,.17)  to[out=0,in=180] (2,.1)
    (0,0.10) arc (90:-70:.1) to[out=0,in=180]  (1,-.17)
    to[out=0,in=180] (2,-.1) arc (-90:90:.1) ;
\draw [draw=black,fill=red!90!black,shift={(3,1)},thick,rotate=20]
    (0,0) circle (.10)
    (0,0.10) to[out=0,in=180]  (1,.17)  to[out=0,in=180] (2,.1)
    (0,0.10) arc (90:-70:.1) to[out=0,in=180]  (1,-.17)
    to[out=0,in=180] (2,-.1) arc (-90:90:.1) ;
\draw [draw=black,fill=red!90!black,shift={(4.5,1)},thick,rotate=20]
    (0,0) circle (.10)
    (0,0.10) to[out=0,in=180]  (1,.17)  to[out=0,in=180] (2,.1)
    (0,0.10) arc (90:-70:.1) to[out=0,in=180]  (1,-.17)
    to[out=0,in=180] (2,-.1) arc (-90:90:.1) ;

\draw [draw=black,fill=red!90!black,shift={(0,2)},thick,rotate=20]
    (0,0) circle (.10)
    (0,0.10) to[out=0,in=180]  (1,.17)  to[out=0,in=180] (2,.1)
    (0,0.10) arc (90:-70:.1) to[out=0,in=180]  (1,-.17)
    to[out=0,in=180] (2,-.1) arc (-90:90:.1) ;
\draw [draw=black,fill=red!90!black,shift={(1.5,2)},thick,rotate=20]
    (0,0) circle (.10)
    (0,0.10) to[out=0,in=180]  (1,.17)  to[out=0,in=180] (2,.1)
    (0,0.10) arc (90:-70:.1) to[out=0,in=180]  (1,-.17)
    to[out=0,in=180] (2,-.1) arc (-90:90:.1) ;
\draw [draw=black,fill=red!90!black,shift={(3,2)},thick,rotate=20]
    (0,0) circle (.10)
    (0,0.10) to[out=0,in=180]  (1,.17)  to[out=0,in=180] (2,.1)
    (0,0.10) arc (90:-70:.1) to[out=0,in=180]  (1,-.17)
    to[out=0,in=180] (2,-.1) arc (-90:90:.1) ;
\draw [draw=black,fill=red!90!black,shift={(4.5,2)},thick,rotate=20]
    (0,0) circle (.10)
    (0,0.10) to[out=0,in=180]  (1,.17)  to[out=0,in=180] (2,.1)
    (0,0.10) arc (90:-70:.1) to[out=0,in=180]  (1,-.17)
    to[out=0,in=180] (2,-.1) arc (-90:90:.1) ;

\draw [draw=black,fill=red!90!black,shift={(0,3)},thick,rotate=20]
    (0,0) circle (.10)
    (0,0.10) to[out=0,in=180]  (1,.17)  to[out=0,in=180] (2,.1)
    (0,0.10) arc (90:-70:.1) to[out=0,in=180]  (1,-.17)
    to[out=0,in=180] (2,-.1) arc (-90:90:.1) ;
\draw [draw=black,fill=red!90!black,shift={(1.5,3)},thick,rotate=20]
    (0,0) circle (.10)
    (0,0.10) to[out=0,in=180]  (1,.17)  to[out=0,in=180] (2,.1)
    (0,0.10) arc (90:-70:.1) to[out=0,in=180]  (1,-.17)
    to[out=0,in=180] (2,-.1) arc (-90:90:.1) ;
\draw [draw=black,fill=red!90!black,shift={(3,3)},thick,rotate=20]
    (0,0) circle (.10)
    (0,0.10) to[out=0,in=180]  (1,.17)  to[out=0,in=180] (2,.1)
    (0,0.10) arc (90:-70:.1) to[out=0,in=180]  (1,-.17)
    to[out=0,in=180] (2,-.1) arc (-90:90:.1) ;
\draw [draw=black,fill=red!90!black,shift={(4.5,3)},thick,rotate=20]
    (0,0) circle (.10)
    (0,0.10) to[out=0,in=180]  (1,.17)  to[out=0,in=180] (2,.1)
    (0,0.10) arc (90:-70:.1) to[out=0,in=180]  (1,-.17)
    to[out=0,in=180] (2,-.1) arc (-90:90:.1) ;

%
\filldraw [fill=black!30!blue,fill opacity=0.2,draw=black,very thick,line join=bevel]
   (-0.3,3.2) -- ++(20:2) -- ++(0:5.1)  -- ++(20:-2) -- cycle ;
\filldraw [fill=black!30!blue,fill opacity=0.2,draw=black,very thick,line join=bevel]
   (4.8,-0.2) -- ++(20:2) -- ++(90:3.4)  -- ++(20:-2) -- cycle ;
\filldraw [fill=black!60!white,draw=black,fill opacity=0.2,very thick,line join=bevel]
   (-0.3,-0.2) rectangle (4.8,3.2) ;

\draw [->,thick] (-0.5,0) +(0,.8) node[left] {\large $y$} +(0,0) -- +(0,1) ;
\draw [->,thick] (0,-0.5) +(2.0,0) node[right] {\large $x$} +(0,0) -- +(2,0) ;
\draw [->,thick] (0,0.3) +(30:.95) node {\large $z$} +(0,0) -- +(20:1.5) ;

\draw [draw=black,very thick] (4.8,-.2)  -- ++(90:3.4) -- ++(20:2) ;
\draw [draw=black,very thick] (4.8,-.2)  -- ++(90:3.4) -- ++(0:-5.1) ;

\end{tikzpicture}}
\caption{In our proposed experimental system,
 the wire medium will not be in free space 
 but will be confined by metallic walls.
Here we represent this in two ways, 
 each containing a finite array of wires with varying radii.
First, 
 we confine the array in a rectangular \emph{waveguide}
 with metal side-walls (in blue-gray), 
 and use periodic boundary conditions 
 to treat wires of infinite length.
Second, 
 we add metallic end-walls (grey) perpendicular to the wires
 to change the waveguide into
 a closed box or \emph{cavity}.
In either case 
 we can do our numerical computations 
 for only one period of variation in the wire radii.
It is very important to note that this 
 variation only corresponds to \emph{half} a period of the electric field.}
\label{fig_wires_box}
\end{figure}
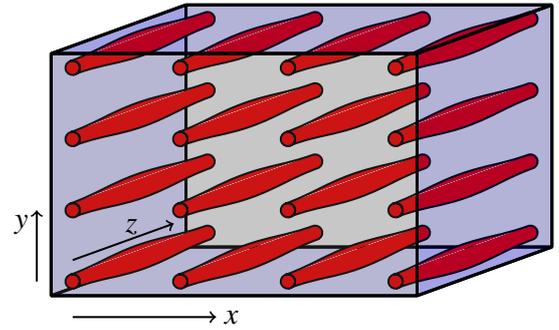

Such field profiling has a variety of potential uses.
 The idea has been implemented by means of
 harmonic synthesis \cite{Chan-HLKLLPP-2011s,Cox-PSLK-2012ol},
 and suggested
 in the context of
 nonlinearity-induced carrier shocking
 \cite{Ward-B-2003prl,Kinsler-RTN-2007pre,Panagiotopoulos-WKM-2015josab}.
Such shaped waveforms have been suggested as a means of
 enhancing ionisation
 in high harmonic generation \cite{Persson-SCB-2006pra,Radnor-CKN-2008pra}.
Other shapes could find uses,
 such as fields with locally high gradients
 but without a large peak --
 minimising nonlinear effects,
 or fields with pronounced peaks
 and low amplitude elsewhere to improve signal to noise ratios.
In particular,
 we are interested in accelerator applications
 where the field profile might be used for electron bunch shaping,
 or as part of a laser wakefield accelerator
 \cite{Piot-SPR-2011prstab,Albert-TMBCFLNVN-2014ppcf}.
This requires us to improve on our initial idealised simulations
 and incorporate experimental features
 such as more realistic material choices
 and a supporting waveguide systems.

In this article we show that the field profiling
 of a longitudinal wave predicted in \cite{Boyd-GKL-2018oe-tbwire}
 can be numerically reproduced in an experimentally plausible system. 
In \cite{Boyd-GKL-2018oe-tbwire}
 we considered an infinite periodic array of infinite wires
 as shown in figure \ref{fig_wires},
  where the wire radius varied from  $0.2\mm$ to $0.4\mm$
 with a relative permittivity of $\epsilon = 1600$.  
As longitudinal EM waves are known to exist
 in guided wave systems we consider a $4 \times 4$ array of finite-wires, 
 a single period long, 
 and contained in a metallic waveguide closed at both ends, 
 as depicted in figure \ref{fig_wires_box}. 
In the simulations reported here, 
 our wire parameters 
 are physically easier to realise
 than those used in \cite{Boyd-GKL-2018oe-tbwire}, 
 being based on a wire radius that varies from $0.5\mm$ to $0.7\mm$,
 with a relative permittivity of $\epsilon = 50$. 
The simulation of this system
 was undertaken using the frequency domain solver
 of the commercial software CST Microwave Studio.
As in \cite{Boyd-GKL-2018oe-tbwire}
 we concentrate on sculpting the electric field profile
 based on the periodic Mathieu function 

\begin{align}
E_z(z) = \text{MathieuCE}\Big(1,0.8,\frac{z L}{2\pi}\Big)
\label{Intro_Mathieu}
\end{align}
where $L$ is the field wavelength,
 and the period of variation in the wire radius is $L/2$.
This field profile has a flatter maximum than a sinusoid,
 and a steeper gradient when passing though zero.

\begin{figure}
\centering
\raisebox{0cm}{\figlabelsize{(a)}}
\includegraphics[height=3cm,width=\columnwidth]{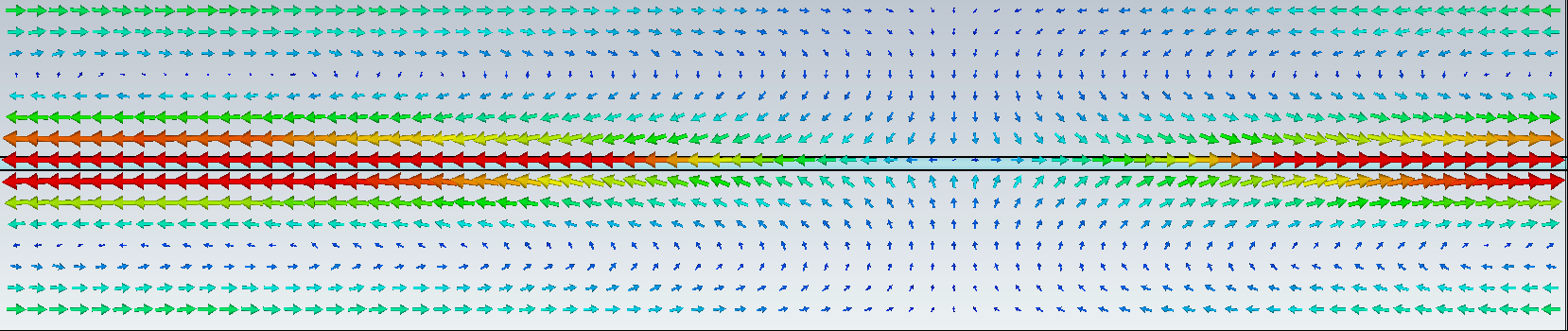}
\\[.5em]
\raisebox{0cm}{\figlabelsize{(b)}}
\includegraphics[height=3cm,width=\columnwidth]{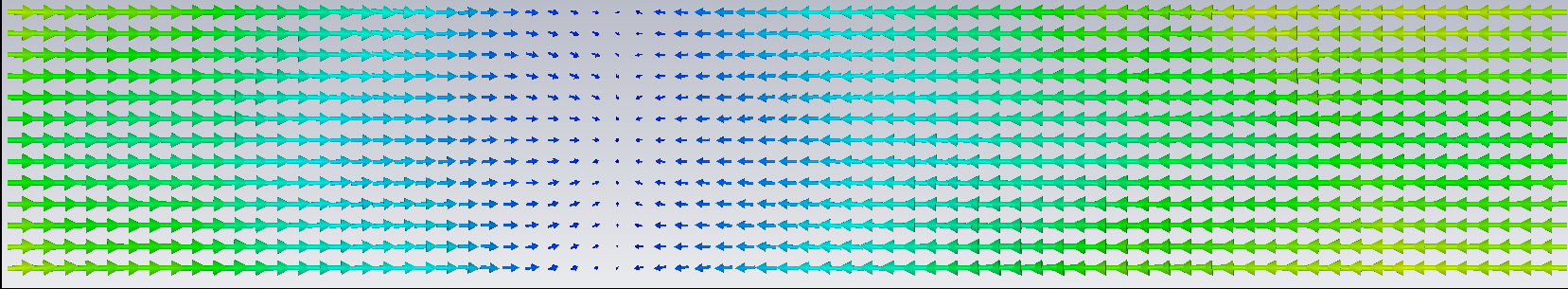}
\\[.5em]
\raisebox{2.35cm}{\figlabelsize{(c)}}
\includegraphics[height=0.5\columnwidth]{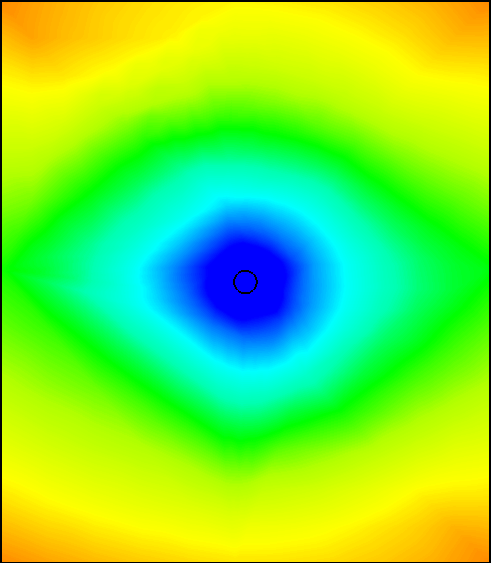}
\\
\qquad
\begin{picture}(4,1)
\put(0,0.5){\includegraphics[width=0.467\columnwidth]{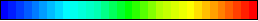}}
\put(0,0){\makebox(0,.2){\small$-1$}}
\put(2,0){{\makebox(0,.2){\small$0$}}}
\put(4,0){\makebox(0,.2){\small$1$}}
\put(.3,0){\small$E_z$ field}
\put(0,0.3){\line(0,1){.2}}
\put(2,0.3){\line(0,1){.2}}
\put(4,0.3){\line(0,1){.2}}
\end{picture}
\caption{Numerical results showing 
 a longitudinal mode 
 in a infinite array of uniform infinitely long wires;
 as depicted in figure \ref{fig_wires}(a). 
We show a vector plot the electric field in the $(y,z)$ plane
 (a) cut through the wires
 and 
 (b) cut between the wires. 
In (c) we show the longitudinal component $E_z$ of the field
 in the $(x,y)$ plane.
Observe there is a region about the wire, 
 roughly circular in shape, 
 where this longitudinal component is approximately zero.}
\label{fig_inf_Unif_Long}
\end{figure}
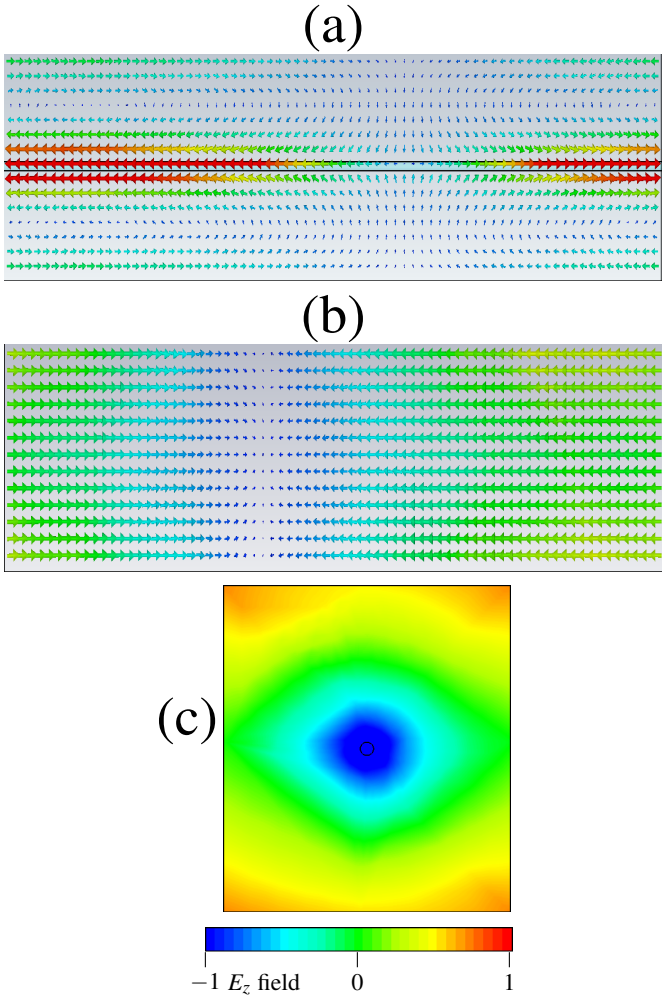

\begin{figure}
\centering
\resizebox{\FIGWIDTH\columnwidth}{!}{
\begin{picture}(5.3,5.5)(-.3,-.5)
\put(0,0){\includegraphics[height=5cm]{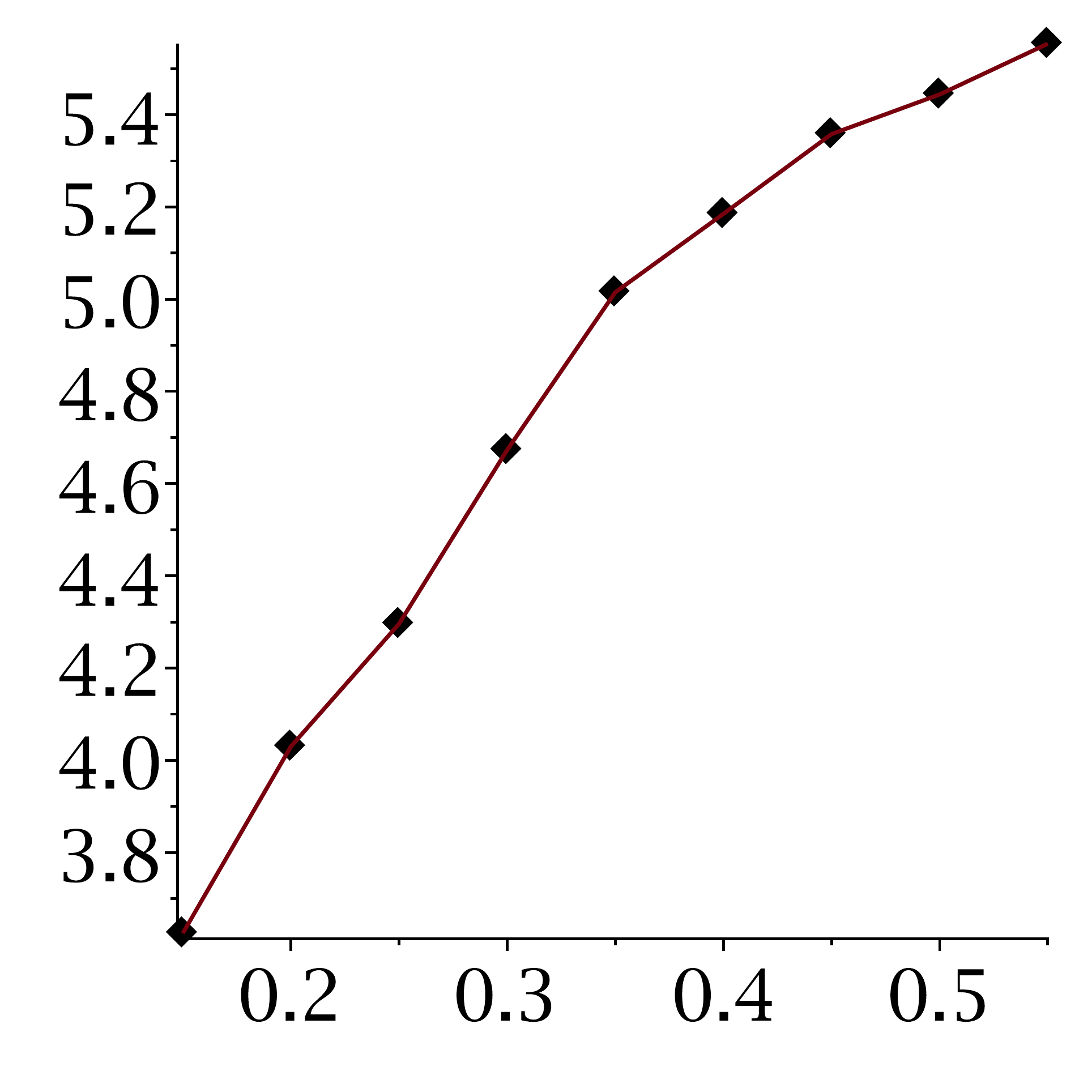}}
\put(-.3,01){\rotatebox{90}{field zero (mm)}}
\put(1,-.2){{wire radius (mm)}}
\end{picture}}
\caption{How the radius of the wire
 affects the distance from the wire 
 at which the longitudinal field  $E_z$ passes through zero.
This result was calculated using periodic boundary conditions
 for an infinite rectangular array
 of uniform and infinitely long wires.
}
\label{fig_inf_Unif_radii_vs_zero}
\end{figure}

\section{Modulated wire media}

This work is based on the understanding of wire media,
 a class of metamaterials consisting of
 a regular (rectangular)  array of parallel wires or rods,
 whose radius is small compared to their spacing
 \cite{Belov-MMNSST-2003prb,Gratus-M-2015jo}.
When the wire radii also vary,
 such three-dimensional inhomogeneous media
 are difficult to analyse in the absence of a simpler model for the system.
Fortunately,
 the fixed-radius case is understood \cite{Belov-MMNSST-2003prb},
 which provides us with a starting point.

Any (fixed-radius) wire medium has an electromagnetic dispersion
 which follows the hydrodynamic Lorentz model,
 with a dependence on a resonance frequency,
 a polariton velocity,
 and a plasma frequency.
In \cite{Boyd-GKL-2018oe-tbwire} we showed that,
 in effect,
 we could assume this held for any thin transverse slice of a wire medium.
Thus,
 by stacking different wire radius slices together,
 to form not a uniform but a varying wire,
 we would construct a medium where
 the plasma frequency changed with position.
This varying plasma frequency then
 leads to a correlated change in
 the effective local refractive index,
 thus --
 according to the electromagnetic wave equation --
 also changing the local curvature of the electric field profile.
Since this curvature control changes the shape of the resulting wave,
 we can calculate how to link any desired field profile
 to a variation in wire radius.

This procedure,
 despite its efficiency,
 is still reliant on numerics.
The analytic calculation of Belov et al. \cite{Belov-MMNSST-2003prb}
 was for infinitely thin wires,
 and so lacked the radius dependence we rely on to control
 their spatially dispersive properties.
However,
 once we complete the step of numerically characterising
 a set of finite radius wires covering a suitable range of sizes,
 we can easily extract the necessary radius dependence of
 the plasma frequency.
Note that although the other hydrodynamic Lorentz parameters
 fitted to the numerical results do change,
 they are much less sensitive,
 and so that variation can be reasonably ignored.

\section{Modulated wires in a supporting waveguide or cavity}

For either an experimental configuration
 or a technological application,
 we would not expect to be able to use an unsupported array of wires
 in free space.
Instead,
 we would need to use a finite array of wires
 in a supporting structure
 such as a metal waveguide or metallic cavity.
Further,
 we would need to arrange the wires within that support
 so that the longitudinal field structure
 and the ability to sculpt the profiles
 is not disrupted or lost.

We choose to model a supporting structure
 that consists of either a metallic waveguide or cavity,
 aligned with a rectangular array of wires with varying radius.
This leaves a number of design parameters to consider,
 (a) the (minimum) number of wires to place in the waveguide or cavity,
 (b) the distance of the waveguide or cavity side-walls parallel to the wires 
 from the nearest wire,
 and,
 for a cavity,
 (c) the separation of the cavity end-walls perpendicular to the wires 
  and their placement
  with respect to the modulation of the wires.

Given that the original simulations \cite{Boyd-GKL-2018oe-tbwire}
 were for an infinite periodic array of wires, 
 we might expect that a large array of many wires would be necessary. 
However,
 we will see below that good results can be achieved
 with only a $2 \times 2$ array; 
 nevertheless in this article we concentrate
 on a $4\times4$ array as we expect it may offer experimental advantages.
Given such a finite array,
 we now need to consider the positioning of the side-walls,
 which would ideally be placed where the longitudinal field is small.
Note that in  figure \ref{fig_inf_Unif_Long}(c)
 we see there is an approximately circular region
 around the wire where the (longitudinal) $E_z$ is zero.
However,
 the $E_z = 0$ contour is not only curved in contrast to
 the planar side-walls of our waveguide or cavity,
 but its distance from the wire varies with the wire radius
 as shown on figure \ref{fig_inf_Unif_radii_vs_zero}.
Nevertheless,
 a reasonable compromise is possible
 since it turns out only requiring $E_z \simeq 0$ is sufficient.

\begin{figure}
\centering
\includegraphics[width=\FIGWIDTH\columnwidth]{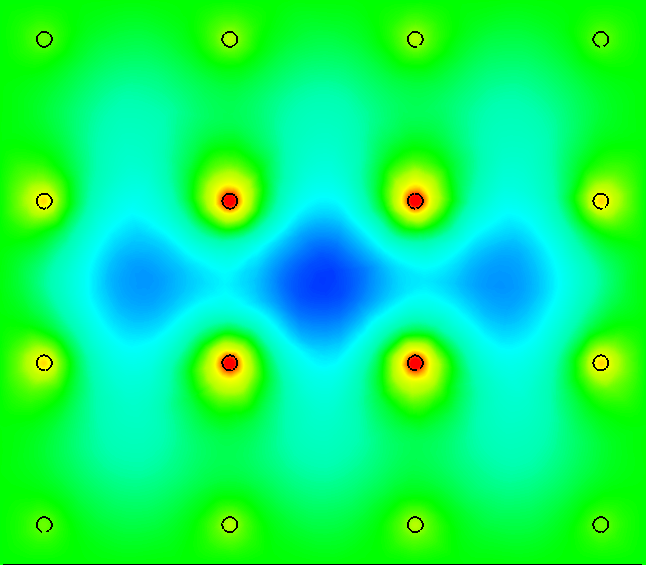}
\\
\begin{picture}(4,1)
\put(0,0.5){\includegraphics[width=0.465\columnwidth]{fig-03-CSTColours.png}}
\put(0,0){\makebox(0,.2){\small$-1$}}
\put(2,0){{\makebox(0,.2){\small$0$}}}
\put(4,0){\makebox(0,.2){\small$1$}}
\put(.3,0){\small$E_z$ field}
\put(0,0.3){\line(0,1){.2}}
\put(2,0.3){\line(0,1){.2}}
\put(4,0.3){\line(0,1){.2}}
\end{picture}
\caption{Longitudinal $E_z$ component on the $(x,y)$ plane
 cutting across a  $4\times 4$ array of uniform
  infinitely long wires 
 in a waveguide. 
This is the data for our preferred longitudinal mode, 
 which is only one of many.}
\label{fig_4x4_Long}
\end{figure}

\begin{figure}
\centering
\resizebox{\FIGWIDTH\columnwidth}{!}{{
\begin{picture}(5.3,5.5)(-.3,-.5)
\put(0,0){\includegraphics[height=5cm]{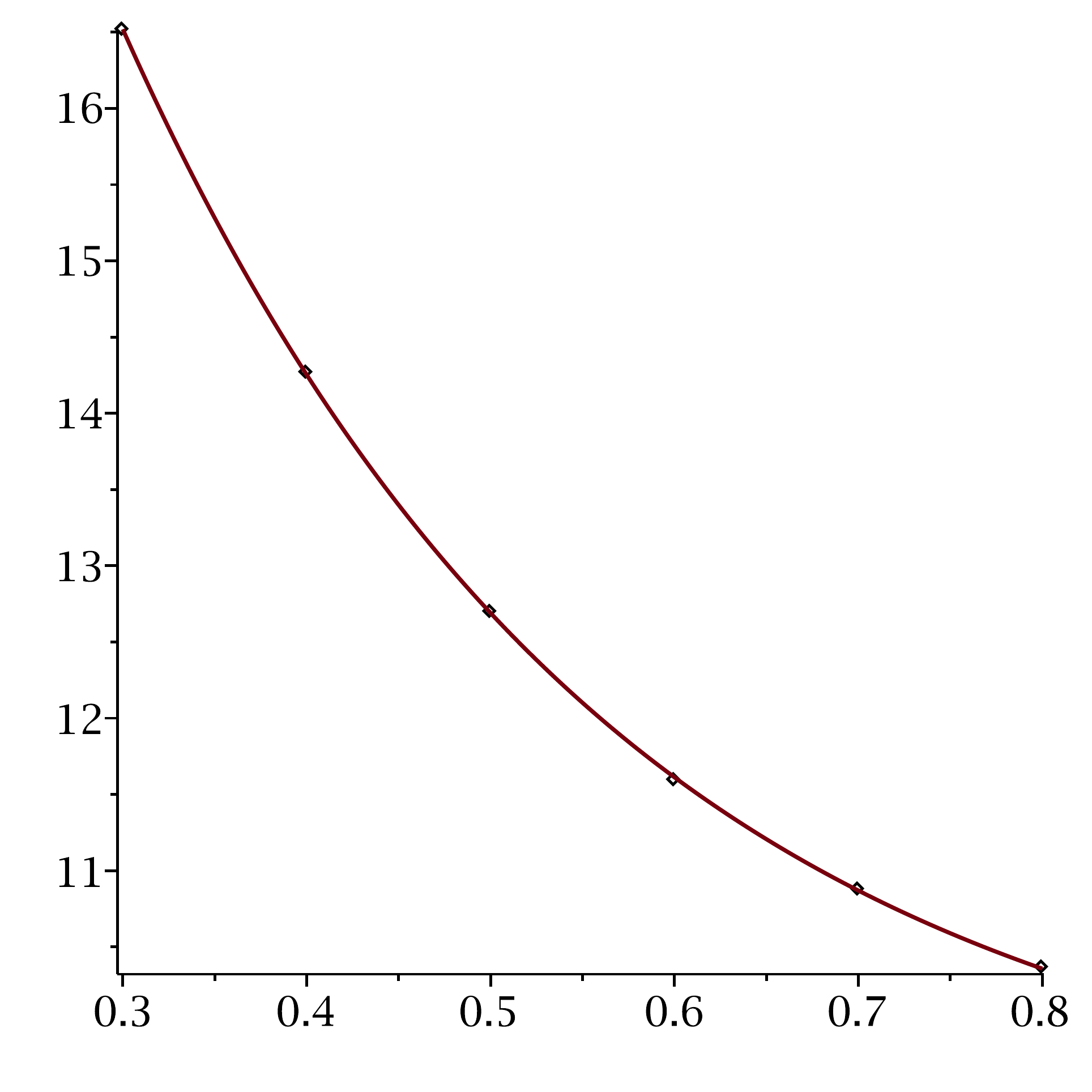}}
\put(-.3,01){\rotatebox{90}{$k_p^2$ \ (GHz${}^2$)}}
\put(1,-.2){{wire radius $r$ (mm)}}
\end{picture}}}
\caption{In order to design the required radius modulation, 
 we need to know the variation of the effective plasma frequency
 of our $4\times 4$ wire array in a waveguide,
 as a function of wire radius, 
 which for our chosen parameters is shown here.
This data is fitted with the
  exponential function $k_p^2(r)=y_0 + A\exp(-r/r_0)$,
 where  $y_0=9.22$, $A=2.23$ and $r_0=2.69$.
}
\label{fig_4x4_Rad_vs_omP}
\end{figure}

\begin{figure}
\centering
\resizebox{\FIGWIDTH\columnwidth}{!}{
\begin{picture}(15.3,4.5)(-.3,.4)
\put(0,0){\includegraphics[height=5cm]{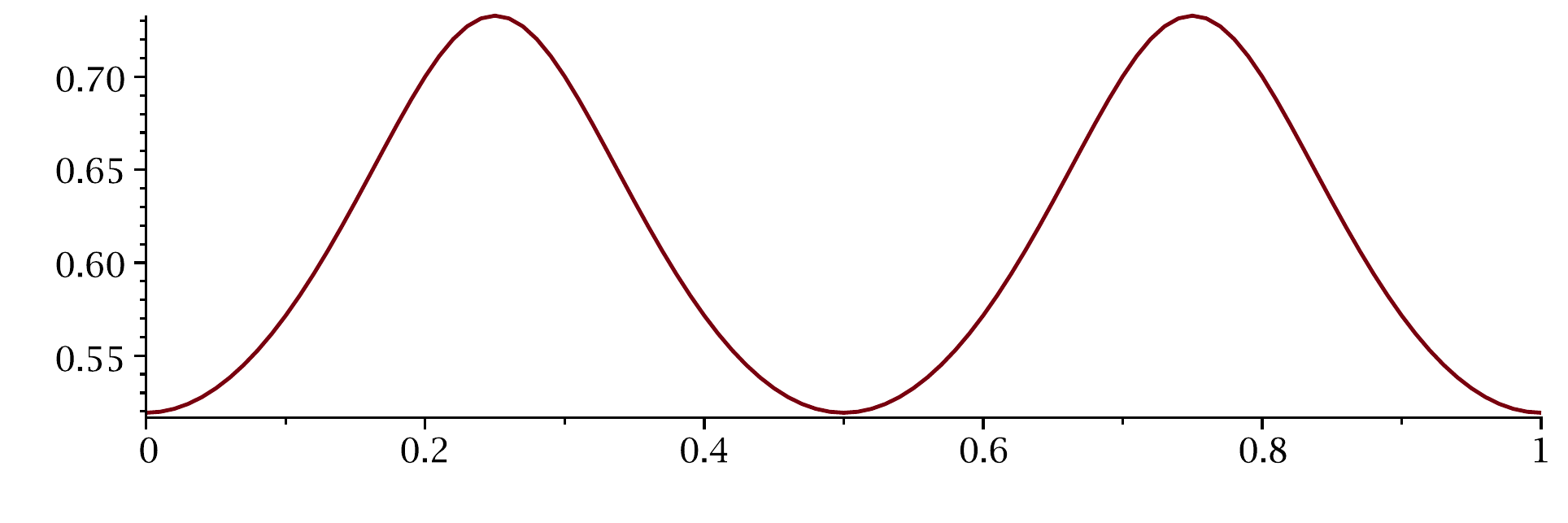}}
\put(.0,01){\rotatebox{90}{wire radius (mm)}}
\put(3,.5){{$z$}}
\put(4.3,0.55){\small$L$}
\put(7,0.55){\small$L$}
\put(9.7,0.55){\small$L$}
\put(12.4,0.55){\small$L$}
\put(14.9,0.55){\small$L$}
\end{picture}}
\caption{By combining our desired field profile
 (the Mathieu function of \eqref{Intro_Mathieu})
 with the plasma frequency response in figure \ref{fig_4x4_Rad_vs_omP}
 we can predict the necessary radius variation
 for our $4\times 4$ wire array in a waveguide.
The result is shown here, 
 with a radius that changes by about $\pm 15$\%
 around its average value.
}
\label{fig_4x4_Rad_profile}
\end{figure}
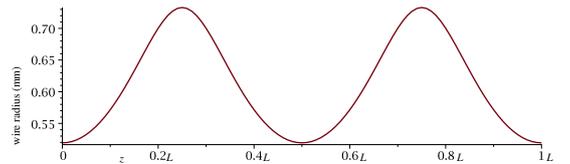

\begin{figure}
\centering
{
\raisebox{2cm}{\figlabelsize{(a)}}
\includegraphics[width=\FIGWIDTH\columnwidth]{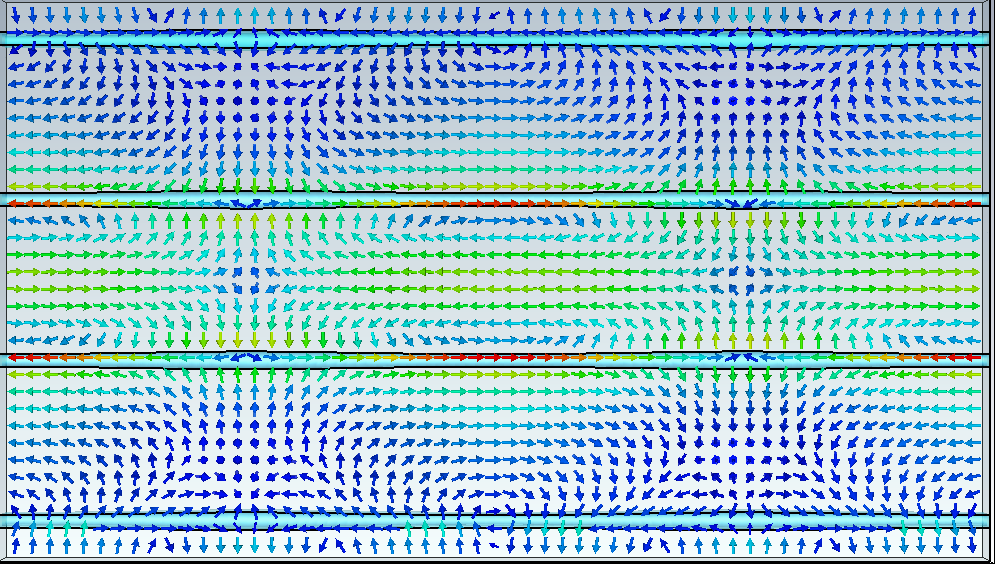}
\qquad
\raisebox{2cm}{\figlabelsize{(b)}}
\includegraphics[width=\FIGWIDTH\columnwidth]{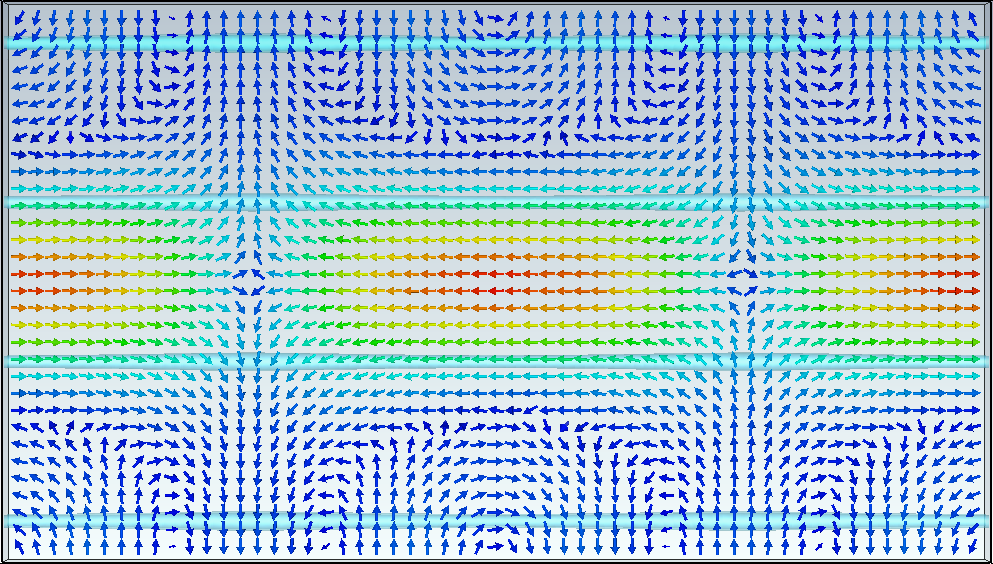}
\\[.5em]
\raisebox{2.75cm}{\figlabelsize{(c)}}
\includegraphics[width=\FIGWIDTH\columnwidth]{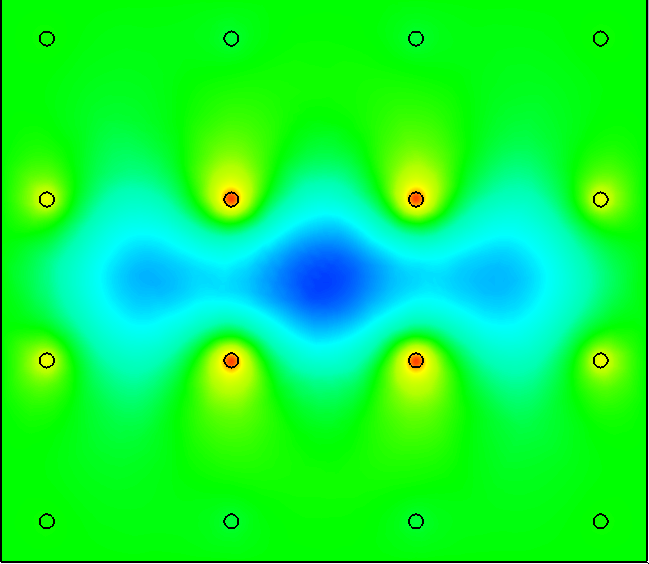}
\\
\qquad
\begin{picture}(4,1)
\put(0,0.5){\includegraphics[width=4cm]{fig-03-CSTColours.png}}
\put(0,0){\makebox(0,.2){\small$-1$}}
\put(2,0){{\makebox(0,.2){\small$0$}}}
\put(4,0){\makebox(0,.2){\small$1$}}
\put(.3,0){\small$E_z$ field}
\put(0,0.3){\line(0,1){.2}}
\put(2,0.3){\line(0,1){.2}}
\put(4,0.3){\line(0,1){.2}}
\end{picture}}
\caption{The field variation in our chosen longitudinal mode
 as present in a $4\times 4$ array of infinitely long wires of varying radii
 in a waveguide. 
The electric field is shown as a vector 
 on the $(y,z)$ plane 
 as (a) cut through the wires,
 and
 (b) cut between the wires. 
In (c), 
 the longitudinal component $E_z$ 
 is shown on the $(x,y)$. }
\label{fig_4x4_Long_Var}
\end{figure}

\begin{figure}
\centering
\resizebox{\FIGWIDTH\columnwidth}{!}{
\begin{picture}(15.3,5.5)(-.3,-.5)
\put(0,0){\includegraphics[height=5cm]{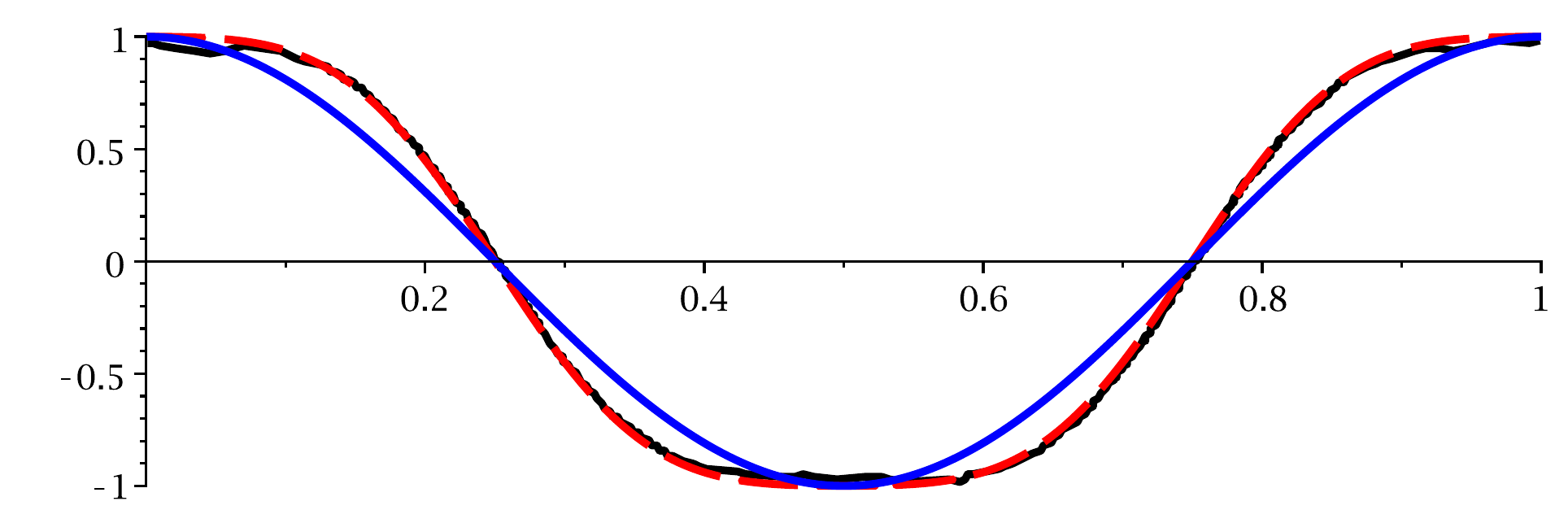}}
\put(.2,2.5){\rotatebox{90}{$E_z$}}
\put(4.3,2){\small$L$}
\put(7,2){\small$L$}
\put(9.7,2){\small$L$}
\put(12.4,2){\small$L$}
\put(14.9,2){\small$L$}
\put(13.5,1.8){$z$}
\end{picture}}
\caption{Field profile shaping in a metallic waveguide
 containing a $4\times 4$ array of infinitely long wires of varying radii
 and with $L=80.61\mm$.
The simulated field profile $E_z$ itself 
 is normalised and shown with solid black lines, 
 and is compared with the ideal Mathieu function profile
 shown as a dashed red curve, 
 with a sinusoid shown in blue as a reference.
In the simulation, 
 CST gave the maximum longitudinal field
 as $3.2810668\times10^7\text{Vm}^{-1}$.}
\label{fig_4x4_profile}
\end{figure}
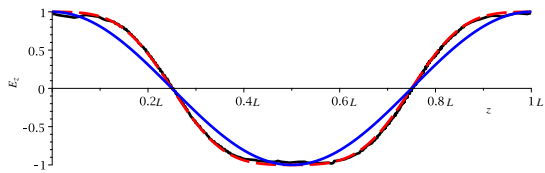


The first step when using our algorithm,
 detailed in \cite{Boyd-GKL-2018oe-tbwire},
 is to check that our proposed waveguide
 containing a $4\times4$ array of uniform
 wires does indeed support longitudinal modes.
As can be seen on figure \ref{fig_4x4_Long}, 
 strong longitudinal fields are indeed present
 in the centre between the inner wires.
Next, 
 we need to calculate the relationship between the wire radius $r$
 and the plasma frequency $k_p^2$, 
 and the results are shown on figure \ref{fig_4x4_Rad_vs_omP} 
 for the $4\times4$ case.
For our chosen field profile, 
 we use this information to calculate the generating radius variation, 
 as seen on figure \ref{fig_4x4_Rad_profile}.
This produced a shaped longitudinal field
 shown in figure \ref{fig_4x4_Long_Var},
 and which is compared with the desired profile on figure  \ref{fig_4x4_profile}.

\def\pSiz{3.65cm}
\def\FIGWIDTHX{0.80}
\def\qSiz{4.25cm}

\begin{figure}
\centering 
\begin{tabular}{r@{\,}c}
\raisebox{\pSiz}{\figlabelsize{(a)}}
\includegraphics[width=\FIGWIDTHX\columnwidth]{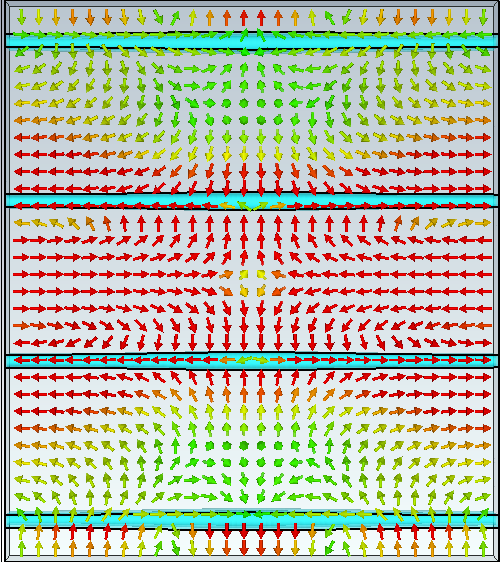}\\
\raisebox{\pSiz}{\figlabelsize{(b)}}
\includegraphics[width=\FIGWIDTHX\columnwidth]{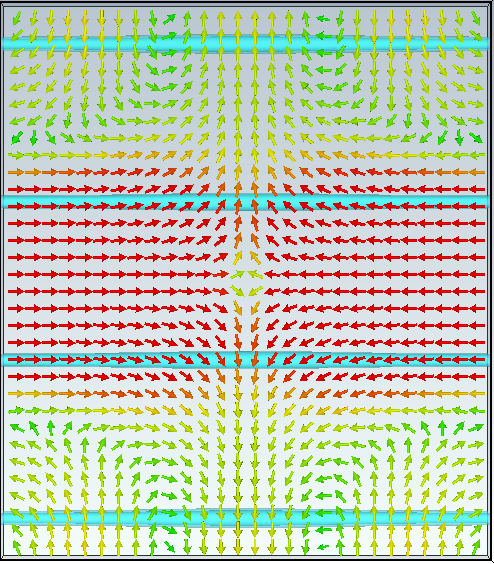}\\
\raisebox{3.0cm}{\figlabelsize{(c)}}
\includegraphics[width=\FIGWIDTHX\columnwidth]{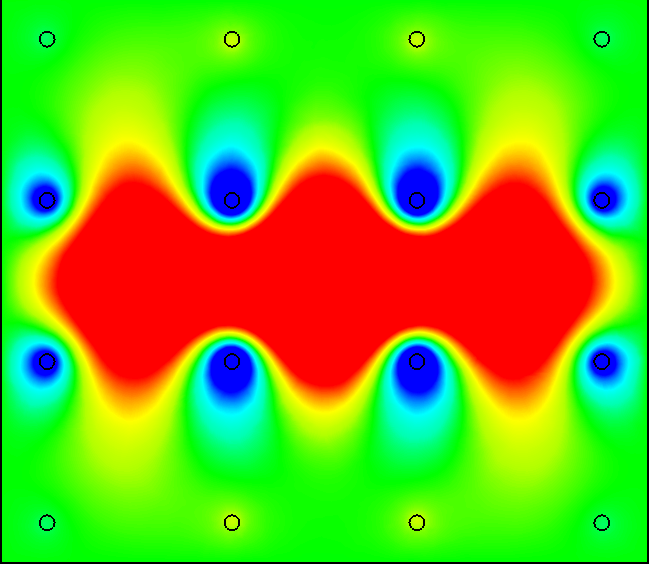}
\\
\begin{picture}(4,1)
\put(0,0.5){\includegraphics[width=4cm]{fig-03-CSTColours.png}}
\put(0,0){\makebox(0,.2){\small$-1$}}
\put(2,0){{\makebox(0,.2){\small$0$}}}
\put(4,0){\makebox(0,.2){\small$1$}}
\put(.3,0){\small$E_z$ field}
\put(0,0.3){\line(0,1){.2}}
\put(2,0.3){\line(0,1){.2}}
\put(4,0.3){\line(0,1){.2}}
\end{picture}
\end{tabular}
\caption{The field variation in our chosen longitudinal mode
 as present in a $4\times 4$ array of wires with varying radii
 inside a metallic cavity,
 as depicted in figure \ref{fig_wires_box}.
The electric field is shown as a vector 
 on the $(y,z)$ plane 
 as (a) cut through the wires,
 and
 (b) cut between the wires. 
In (c), 
 the longitudinal component $E_z$ 
 is shown on the $(x,y)$. }
\label{fig_4x4_Box_Long_Var}
\end{figure}

\begin{figure}
\centering
\resizebox{\FIGWIDTH\columnwidth}{!}{
\begin{picture}(7.3,5.5)(-.3,-.5)
\put(0,0){\includegraphics[height=5cm]{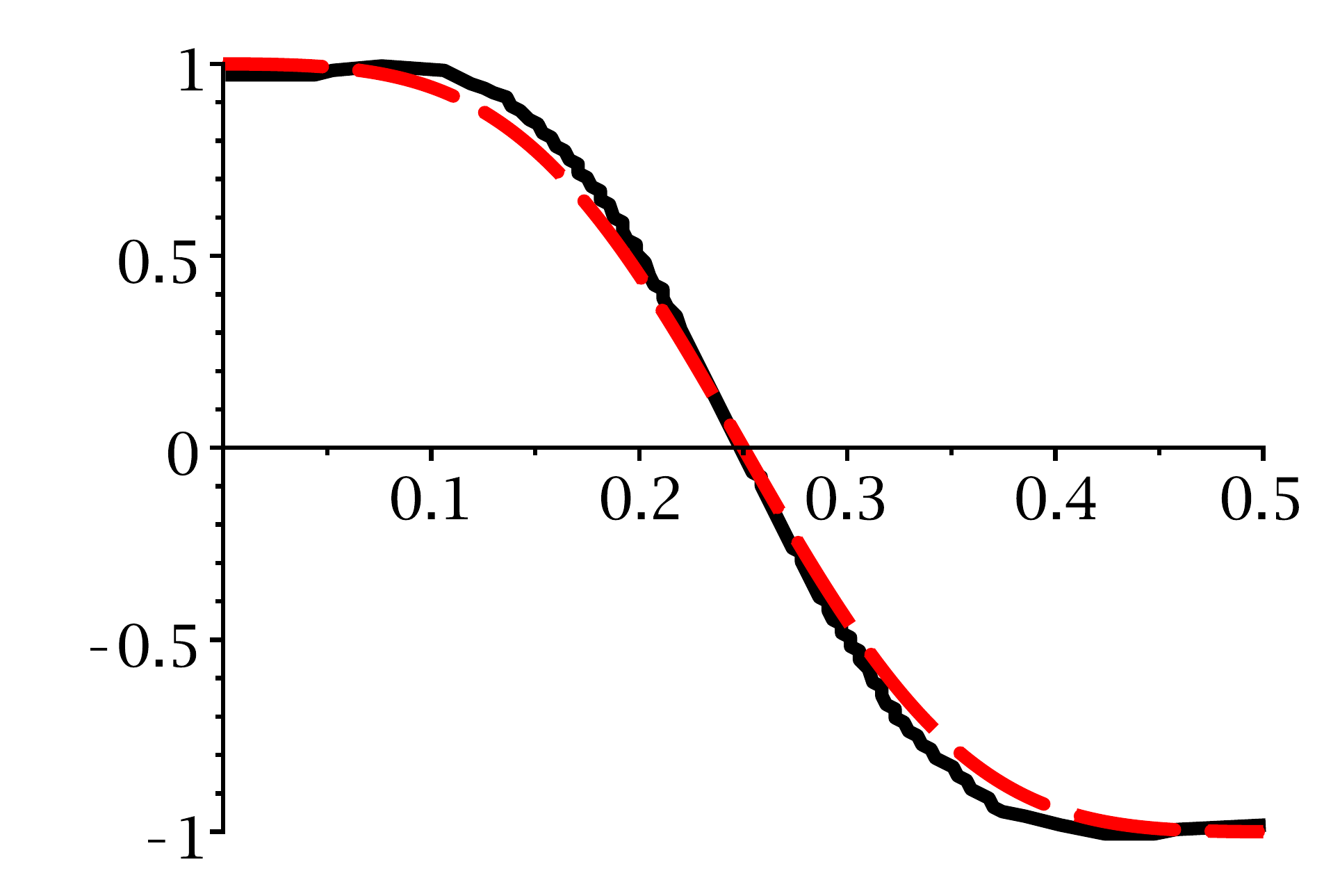}}
\put(.2,2.5){\rotatebox{90}{$E_z$}}
\put(2.6,2.07){\small$L$}
\put(3.8,2.07){\small$L$}
\put(5.0,2.07){\small$L$}
\put(6.2,2.07){\small$L$}
\put(7.3,2.07){\small$L$}
\put(3,1.8){$z$}
\end{picture}}
\caption{Field profile shaping in a metallic waveguide
 containing a $4\times 4$ array of wires with varying radii
 in a metallic cavity, 
 and with $L=80.61\mm$.
The simulated field profile $E_z$ itself 
 is normalised and shown with solid black lines, 
 and is compared with the ideal Mathieu function profile
 shown as a dashed red curve.}
\label{fig_4x4_Box_profile}
\end{figure}
\vspace{1em}

\begin{figure}
\centering
\includegraphics[width=\FIGWIDTH\columnwidth]{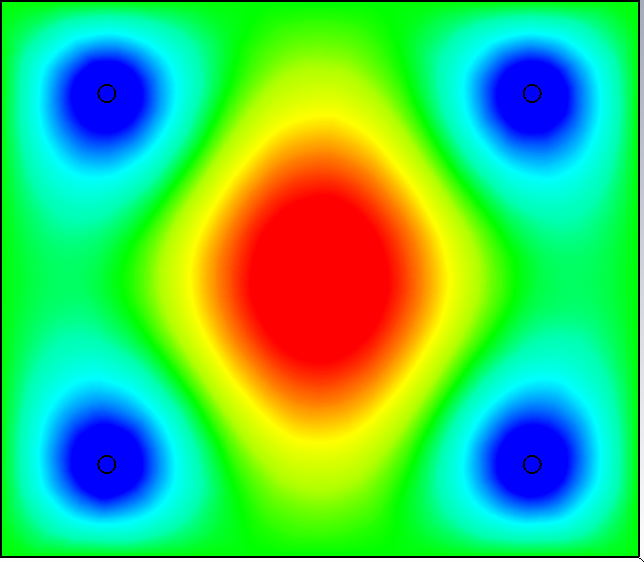}
\\
\begin{picture}(4,1)
\put(0,0.5){\includegraphics[width=4cm]{fig-03-CSTColours.png}}
\put(0,0){\makebox(0,.2){\small$-1$}}
\put(2,0){{\makebox(0,.2){\small$0$}}}
\put(4,0){\makebox(0,.2){\small$1$}}
\put(.3,0){\small$E_z$ field}
\put(0,0.3){\line(0,1){.2}}
\put(2,0.3){\line(0,1){.2}}
\put(4,0.3){\line(0,1){.2}}
\end{picture}
\caption{Field variation of a longitudinal modes
 in a $2\times 2$ array of uniform infinitely long wires
 \textbf{in a waveguide}.
Here the
  longitudinal component $E_z$ in the $(x,y)$ plane is shown.}
\label{fig_2x2_Long}
\end{figure}

\begin{figure}
\centering
\resizebox{\FIGWIDTH\columnwidth}{!}{
\begin{picture}(15.0,5.5)(-.0,-.5)
\put(0,0){\includegraphics[height=5cm]{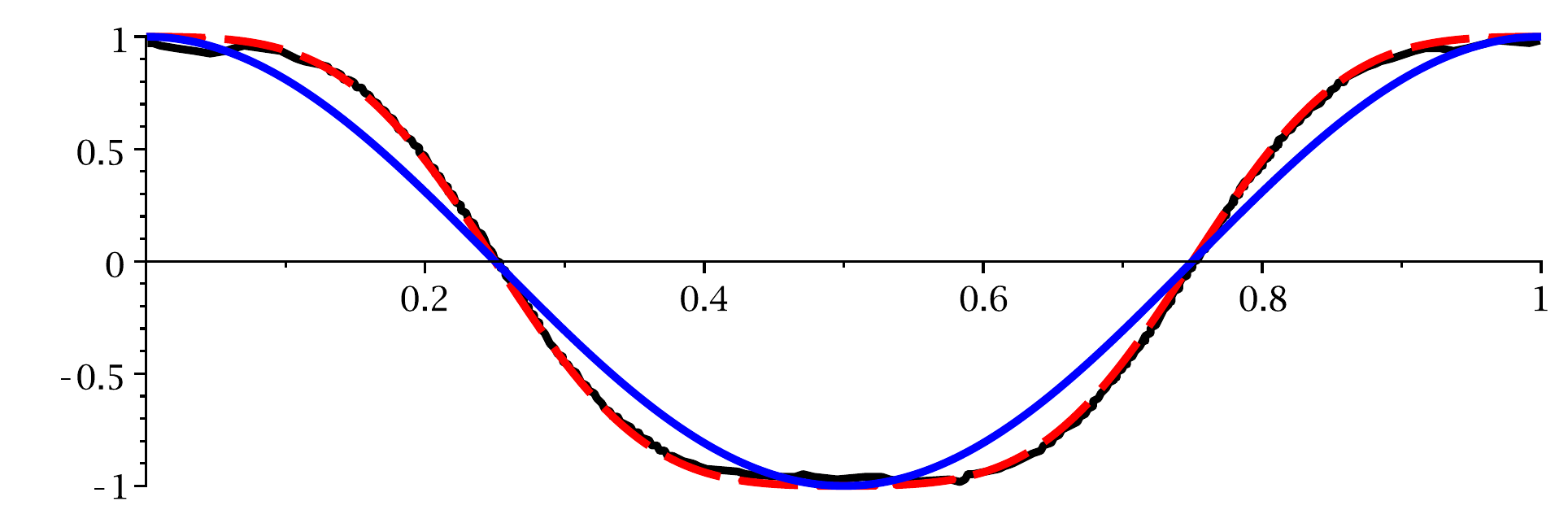}}
\put(.2,2.5){\rotatebox{90}{$E_z$}}
\put(4.3,2){\small$L$}
\put(7,2){\small$L$}
\put(9.7,2){\small$L$}
\put(12.4,2){\small$L$}
\put(14.9,2){\small$L$}
\put(13.5,1.8){$z$}
\end{picture}}
\caption{Profile of
      $E_z$ component for a waveguide with $2\times 2$ infinitely
  long wires of
  varying radii (solid Black) in comparison with Mathieu function
  (dashed red) and sin function (blue). Here $L=127.4\mm$. Both $E_z$
  and the Mathieu function are normalised. However CST gave the maximum
  longitudinal field as $6.03\times10^7\text{Vm}^{-1}$.}
\label{fig_2x2_profile}
\end{figure}
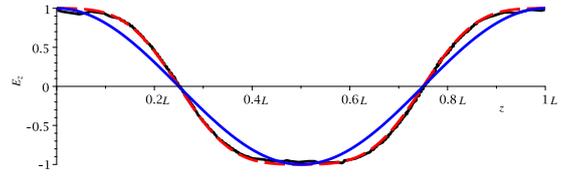

The next step was to consider the case of a cavity 
 rather than the infinite wavguide 
 which requires us to choose where to place
 the perpendicular end-walls.
In this case we made the straightforward decision 
 to place them where the field has a symmetry, 
 i.e. at its maximum and minimum values. 
Thus the cavity contains an exact half wavelength of the field, 
 which is the same as one full period of the wire variation.
We see in figure \ref{fig_4x4_Box_Long_Var}
 that the longitudinal profile of the mode is unaffected
 and from figure \ref{fig_4x4_Box_profile} that
 the simulated field profiles is still in very good agreement
 with the intended Mathieu profile.
We also repeated the above procedure for the simpler cavity
 with only a $2\times2$ wire array,
 as shown on figure \ref{fig_2x2_Long} and \ref{fig_2x2_profile}. 
Even in this reduced situation 
 there was still excellent agreement with the desired electric field profile.

\section{Frequency Isolation}

\begin{figure}
\centering
\resizebox{\FIGWIDTH\columnwidth}{!}{
\includegraphics[height=5cm]{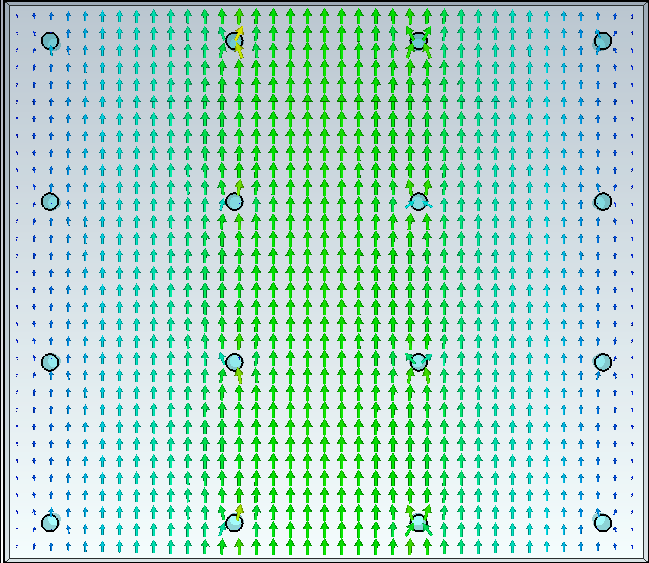}
}
\caption{An example
 showing the electric field vectors on a slice in the $(x,y)$ plane
 perpendicular to the wires.
Most of the field modes near our desired longitudinal modes
 are of this transverse type, 
 and should not be excited by a carefully designed 
 (longitudinal)
 excitation field.}
\label{fig_transverse_mode}
\end{figure}

When attempting to experimentally demonstrate profile shaping
 it is necessary to not merely excite the appropriate mode in the structure, 
 but also excite \emph{only} that mode.
In particular, 
 we need to be able to avoid  exciting the nearby modes
 which may not be longitudinal, 
 nor be correctly shaped.
To assist with this selection, 
 in table \ref{tab_freq} we show 
 an eigenmode analysis for the cavity
 in the vicinity of the index 90 
 ($12.430$GHz) mode we wish to excite.
At first sight this does not look promising, 
 with nearby modes being within 0.3\% 
 (i.e. about $0.04$GHz).
However, 
 we can also see that the nearest modes are transverse 
 (as in figure \ref{fig_transverse_mode})
 rather than longitudinal, 
 so that they should not be strongly excited 
 as long as we only attempt to excite the desired mode
 with a well designed source.
The nearest unwanted longitudinal mode is further away in frequency
 (about  0.9\% or $0.1$GHz), 
 so it need not present too much of a challenge
 if we can constrain the bandwidth of the mode and source sufficiently.

\begin{table}
\begin{tabular}{|l|l|l|l|l|l|l|l|l|l|}
\hline
Mode Number & 
82 & 83 & 84 & 85 & 86 & 87 
\\\hline
Freq (GHz) &
12.088 &
12.098 &
12.158 &
12.159 &
12.175 &
12.250 
\\\hline
Type  &
Long  &
Long  &
Long  &
Trans &
Long  &
Trans 
\\\hline
\end{tabular}
\\[.5em]
\begin{tabular}{|l|l|l|l|l|l|l|l|l|l|}
\hline
Mode Number & 
88 & 
89 &
\textit{\textbf{90}} &
 91    &
 92    &
 93    
\\\hline
Freq (GHz) &
12.274 &
12.388 &
\textit{\textbf{12.430}} &
12.433   &
12.460   &
12.539   
\\\hline
Type  &
Trans &
Trans &
\textit{\textbf{Long}} &
Trans   &
Trans   &
Long    
\\\hline
\end{tabular}
\\[.5em]
\begin{tabular}{|l|l|l|l|l|l|l|l|l|l|}
\hline
Mode Number & 
 94    &
 95    &
 96    &
 97    &
 98    &
 99   
\\\hline
Freq (GHz) &
12.556   &
12.565   &
12.577   &
12.583   &
12.723   &
12.831  
\\\hline
Type  &
Long    &
Long    &
Long    &
Trans   &
Long    &
Trans 
\\\hline
\end{tabular}
\caption{Modes of the modulated wire media
 in a cavity that are
 nearby to our chosen mode (index 90).
They are categorized into either transverse (`Trans')
 or longitudinal modes (`Long'), 
 according to the behaviour of $E_z$.}
\label{tab_freq}
\end{table}


\section{Conclusion}

We have presented evidence that it should be possible 
 to reproduce the field profile shaping results
 of \cite{Boyd-GKL-2018oe-tbwire}
 in an experimentally realisable system.
The system consists of a finite number of wires,
 a single period long,
 inside a metal cavity, 
 and despite the many potential difficulties introduced by the cavity,
 simulations still show that we 
 can modify the profile 
 of the longitudinal electric field at will.
Further, 
 the shaped mode that we intend to excite should be 
 sufficiently distinct from other nearby modes
 to allow the effect to be observed in practise.
Nevertheless, 
 more work is needed to improve this separation,
 either by identifying 
 better methods of frequency isolation
 or by exploiting the properties of the nearby modes.

Despite these promising predictions, 
 there are other practical challenges that need to be overcome.
We know that wires with a relative permittivity of $\epsilon = 50$
 can be produced from a composite blend of $TiO$ particles and plasticiser, 
 but thin wires with radii between $0.5\mm$ and $0.7\mm$
 may be impractical to work with, 
 even via controlled extrusion. 
If such manufacturing constraints cannot be overcome, 
 we can easily redesign the system 
 to use a larger cavity, 
 work at a lower frequency, 
 and need lower permittivity wires.
Of course such adjustment will need to be led by 
 the actual experimental design
 and lessons learned during its implementation and testing.

Beyond the challenges involved in construction of the 
 field profile shaping apparatus, 
 there is also the question of what is the best method
 for exciting the longitudinal modes, 
 and what the best way to measure them.
One approach would be to load a sample of our wire array media
 into a section of a larger waveguide.
In this way the naturally occurring longitudinal wave of the bare waveguide 
 could couple naturally to the longitudinal (and shaped)
 wave in the loaded section, 
 with minimal effect on the unwanted -- 
 but nearby in frequency --
 transverse modes.
For measurement, 
 a probe attached to a spectrum analyser 
 can be used to map the internal field profile, 
 or, 
 alternatively, 
 a perturbation bead-pull approach could be used. 

In summary, 
 we believe our simulations provide sufficient evidence
 that our field profile shaping method will work in practise; 
 justifying our intent to proceed with designing and building
 an experiment to verify this.
Nevertheless, 
 the frequency domain simulations done here 
 could still be extended upon -- 
 notably we would like to move forward to time domain simulation
 to provide more information, 
 such as to how the field profile will build up in the wire array media
 when the excitation field is turned on.

\section{Acknowledgement}
TB, JG, PK, and RL are grateful for the support provided by STFC 
 (the Cockcroft Institute 
   ST/P002056/1), 
 and JG and PK are grateful for the support provided by 
  EPSRC 
  (Alpha-X project EP/N028694/1).

\bibliography{bibexport}

\begin{thebibliography}{14}
\expandafter\ifx\csname natexlab\endcsname\relax\def\natexlab#1{#1}\fi
\expandafter\ifx\csname bibnamefont\endcsname\relax
  \def\bibnamefont#1{#1}\fi
\expandafter\ifx\csname bibfnamefont\endcsname\relax
  \def\bibfnamefont#1{#1}\fi
\expandafter\ifx\csname citenamefont\endcsname\relax
  \def\citenamefont#1{#1}\fi
\expandafter\ifx\csname url\endcsname\relax
  \def\url#1{\texttt{#1}}\fi
\expandafter\ifx\csname urlprefix\endcsname\relax\def\urlprefix{URL }\fi
\providecommand{\bibinfo}[2]{#2}
\providecommand{\eprint}[2][]{\url{#2}}

\bibitem[{\citenamefont{Gratus et~al.}(2017{\natexlab{a}})\citenamefont{Gratus,
  Kinsler, Letizia, and Boyd}}]{Gratus-KLB-2017apa-malaga}
\bibinfo{author}{\bibfnamefont{J.}~\bibnamefont{Gratus}},
  \bibinfo{author}{\bibfnamefont{P.}~\bibnamefont{Kinsler}},
  \bibinfo{author}{\bibfnamefont{R.}~\bibnamefont{Letizia}}, \bibnamefont{and}
  \bibinfo{author}{\bibfnamefont{T.}~\bibnamefont{Boyd}},\\
  \bibinfo{journal}{Appl. Phys. A} \textbf{\bibinfo{volume}{123}},
  \bibinfo{pages}{108} (\bibinfo{year}{2017}{\natexlab{a}}),\\
  \XDOI{10.1007/s00339-016-0649-8}.

\bibitem[{\citenamefont{Gratus et~al.}(2017{\natexlab{b}})\citenamefont{Gratus,
  Kinsler, Letizia, and Boyd}}]{Gratus-KLB-2017jpc}
\bibinfo{author}{\bibfnamefont{J.}~\bibnamefont{Gratus}},
  \bibinfo{author}{\bibfnamefont{P.}~\bibnamefont{Kinsler}},
  \bibinfo{author}{\bibfnamefont{R.}~\bibnamefont{Letizia}}, \bibnamefont{and}
  \bibinfo{author}{\bibfnamefont{T.}~\bibnamefont{Boyd}}, \\
\bibinfo{journal}{J.
  Phys. Commun.} \textbf{\bibinfo{volume}{1}}, \bibinfo{pages}{025003}
  (\bibinfo{year}{2017}{\natexlab{b}}),\\
  \XDOI{10.1088/2399-6528/aa81bb}.

\bibitem[{\citenamefont{Boyd et~al.}(2018)\citenamefont{Boyd, Gratus, Kinsler,
  and Letizia}}]{Boyd-GKL-2018oe-tbwire}
\bibinfo{author}{\bibfnamefont{T.}~\bibnamefont{Boyd}},
  \bibinfo{author}{\bibfnamefont{J.}~\bibnamefont{Gratus}},
  \bibinfo{author}{\bibfnamefont{P.}~\bibnamefont{Kinsler}}, \bibnamefont{and}
  \bibinfo{author}{\bibfnamefont{R.}~\bibnamefont{Letizia}},\\
  \bibinfo{journal}{Opt. Express} \textbf{\bibinfo{volume}{26}},
  \bibinfo{pages}{2478} (\bibinfo{year}{2018}),\\
  \XDOI{10.1364/OE.26.002478}.

\bibitem[{\citenamefont{Chan et~al.}(2011)\citenamefont{Chan, Hsieh, Liang,
  Kung, Lee, Lai, Pan, and Peng}}]{Chan-HLKLLPP-2011s}
\bibinfo{author}{\bibfnamefont{H.-S.} \bibnamefont{Chan}},
  \bibinfo{author}{\bibfnamefont{Z.-M.} \bibnamefont{Hsieh}},
  \bibinfo{author}{\bibfnamefont{W.-H.} \bibnamefont{Liang}},
  \bibinfo{author}{\bibfnamefont{A.~H.} \bibnamefont{Kung}},
  \bibinfo{author}{\bibfnamefont{C.-K.} \bibnamefont{Lee}},
  \bibinfo{author}{\bibfnamefont{C.-J.} \bibnamefont{Lai}},
  \bibinfo{author}{\bibfnamefont{R.-P.} \bibnamefont{Pan}}, \bibnamefont{and}
  \bibinfo{author}{\bibfnamefont{L.-H.} \bibnamefont{Peng}},\\
  \bibinfo{journal}{Science} \textbf{\bibinfo{volume}{331}},
  \bibinfo{pages}{1165} (\bibinfo{year}{2011}),\\
  \XDOI{10.1126/science.1198397}.

\bibitem[{\citenamefont{Cox et~al.}(2012)\citenamefont{Cox, Putnam, Sell,
  Leitenstorfer, and K\"artner}}]{Cox-PSLK-2012ol}
\bibinfo{author}{\bibfnamefont{J.~A.} \bibnamefont{Cox}},
  \bibinfo{author}{\bibfnamefont{W.~P.} \bibnamefont{Putnam}},
  \bibinfo{author}{\bibfnamefont{A.}~\bibnamefont{Sell}},
  \bibinfo{author}{\bibfnamefont{A.}~\bibnamefont{Leitenstorfer}},
 \bibinfo{author}{\bibfnamefont{F.~X.}
  \bibnamefont{K\"artner}}, \\
\bibinfo{journal}{Opt. Lett.}
  \textbf{\bibinfo{volume}{37}}, \bibinfo{pages}{3579} (\bibinfo{year}{2012}),\\
  \XDOI{10.1364/OL.37.003579}.

\bibitem[{\citenamefont{Ward and Berge}(2003)}]{Ward-B-2003prl}
\bibinfo{author}{\bibfnamefont{H.}~\bibnamefont{Ward}} \bibnamefont{and}
  \bibinfo{author}{\bibfnamefont{L.}~\bibnamefont{Berge}},\\
  \bibinfo{journal}{Phys. Rev. Lett.} \textbf{\bibinfo{volume}{90}},
  \bibinfo{pages}{053901} (\bibinfo{year}{2003}),\\
  \XDOI{10.1103/PhysRevLett.90.053901}.

\bibitem[{\citenamefont{Kinsler et~al.}(2007)\citenamefont{Kinsler, Radnor,
  Tyrrell, and New}}]{Kinsler-RTN-2007pre}
\bibinfo{author}{\bibfnamefont{P.}~\bibnamefont{Kinsler}},
  \bibinfo{author}{\bibfnamefont{S.~B.~P.} \bibnamefont{Radnor}},
  \bibinfo{author}{\bibfnamefont{J.~C.~A.} \bibnamefont{Tyrrell}},
  \bibnamefont{and} \bibinfo{author}{\bibfnamefont{G.~H.~C.}
  \bibnamefont{New}}, \\
\bibinfo{journal}{Phys. Rev. E}
  \textbf{\bibinfo{volume}{75}}, \bibinfo{pages}{066603}
  (\bibinfo{year}{2007}),\\
  \XDOI{10.1103/PhysRevE.75.066603}, \\
  \XARXIV{0704.1212}.

\bibitem[{\citenamefont{Panagiotopoulos
  et~al.}(2015)\citenamefont{Panagiotopoulos, Whalen, Kolesik, and
  Moloney}}]{Panagiotopoulos-WKM-2015josab}
\bibinfo{author}{\bibfnamefont{P.}~\bibnamefont{Panagiotopoulos}},
  \bibinfo{author}{\bibfnamefont{P.}~\bibnamefont{Whalen}},
  \bibinfo{author}{\bibfnamefont{M.}~\bibnamefont{Kolesik}}, \bibnamefont{and}
  \bibinfo{author}{\bibfnamefont{J.~V.} \bibnamefont{Moloney}},\\
  \bibinfo{journal}{J. Opt. Soc. Am. B} \textbf{\bibinfo{volume}{32}},
  \bibinfo{pages}{1718} (\bibinfo{year}{2015}),\\
  \XDOI{10.1364/JOSAB.32.001718}.

\bibitem[{\citenamefont{Persson et~al.}(2006)\citenamefont{Persson, Schiessl,
  Scrinzi, and Burgdorfer}}]{Persson-SCB-2006pra}
\bibinfo{author}{\bibfnamefont{E.}~\bibnamefont{Persson}},
  \bibinfo{author}{\bibfnamefont{K.}~\bibnamefont{Schiessl}},
  \bibinfo{author}{\bibfnamefont{A.}~\bibnamefont{Scrinzi}}, \bibnamefont{and}
  \bibinfo{author}{\bibfnamefont{J.}~\bibnamefont{Burgdorfer}},\\
  \bibinfo{journal}{Phys. Rev. A} \textbf{\bibinfo{volume}{74}},
  \bibinfo{pages}{013818} (\bibinfo{year}{2006}),\\
  \XDOI{10.1103/PhysRevA.74.013818}.

\bibitem[{\citenamefont{Radnor et~al.}(2008)\citenamefont{Radnor, Chipperfield,
  Kinsler, and New}}]{Radnor-CKN-2008pra}
\bibinfo{author}{\bibfnamefont{S.~B.~P.} \bibnamefont{Radnor}},
  \bibinfo{author}{\bibfnamefont{L.~E.} \bibnamefont{Chipperfield}},
  \bibinfo{author}{\bibfnamefont{P.}~\bibnamefont{Kinsler}}, 
  \bibinfo{author}{\bibfnamefont{G.~H.~C.} \bibnamefont{New}},\\
  \bibinfo{journal}{Phys. Rev. A} \textbf{\bibinfo{volume}{77}},
  \bibinfo{pages}{033806} (\bibinfo{year}{2008}),\\
  \XDOI{10.1103/PhysRevA.77.033806},\\
  \XARXIV{0803.3597}.

\bibitem[{\citenamefont{Piot et~al.}(2011)\citenamefont{Piot, Sun, Power, and
  Rihaoui}}]{Piot-SPR-2011prstab}
\bibinfo{author}{\bibfnamefont{P.}~\bibnamefont{Piot}},
  \bibinfo{author}{\bibfnamefont{Y.~E.} \bibnamefont{Sun}},
  \bibinfo{author}{\bibfnamefont{J.~G.} \bibnamefont{Power}}, \bibnamefont{and}
  \bibinfo{author}{\bibfnamefont{M.}~\bibnamefont{Rihaoui}},\\
  \bibinfo{journal}{Phys. Rev. ST Accel. Beams} \textbf{\bibinfo{volume}{14}},
  \bibinfo{pages}{022801} (\bibinfo{year}{2011}),\\
  \XDOI{10.1103/PhysRevSTAB.14.022801}.

\bibitem[{\citenamefont{Albert et~al.}(2014)\citenamefont{Albert, Thomas,
  Mangles, Banerjee, Corde, Flacco, Litos, Neely, Vieira, and
  Najmudin}}]{Albert-TMBCFLNVN-2014ppcf}
\bibinfo{author}{\bibfnamefont{F.}~\bibnamefont{Albert}},
  \bibinfo{author}{\bibfnamefont{A.~G.~R.} \bibnamefont{Thomas}},
  \bibinfo{author}{\bibfnamefont{S.~P.~D.} \bibnamefont{Mangles}},
  \bibinfo{author}{\bibfnamefont{S.}~\bibnamefont{Banerjee}},
  \bibinfo{author}{\bibfnamefont{S.}~\bibnamefont{Corde}},
  \bibinfo{author}{\bibfnamefont{A.}~\bibnamefont{Flacco}},
  \bibinfo{author}{\bibfnamefont{M.}~\bibnamefont{Litos}},
  \bibinfo{author}{\bibfnamefont{D.}~\bibnamefont{Neely}},
  \bibinfo{author}{\bibfnamefont{J.}~\bibnamefont{Vieira}}, 
  \bibinfo{author}{\bibfnamefont{Z.}~\bibnamefont{Najmudin}},\\
  \bibinfo{journal}{Plasma Physics and Controlled Fusion}
  \textbf{\bibinfo{volume}{56}}, \bibinfo{pages}{084015}
  (\bibinfo{year}{2014}),\\
  \XDOI{10.1088/0741-3335/56/8/084015}.

\bibitem[{\citenamefont{Belov et~al.}(2003)\citenamefont{Belov, Marqu\'es,
  Maslovski, Nefedov, Silveirinha, Simovski, and
  Tretyakov}}]{Belov-MMNSST-2003prb}
\bibinfo{author}{\bibfnamefont{P.~A.} \bibnamefont{Belov}},
  \bibinfo{author}{\bibfnamefont{R.}~\bibnamefont{Marqu\'es}},
  \bibinfo{author}{\bibfnamefont{S.~I.} \bibnamefont{Maslovski}},
  \bibinfo{author}{\bibfnamefont{I.~S.} \bibnamefont{Nefedov}},
  \bibinfo{author}{\bibfnamefont{M.}~\bibnamefont{Silveirinha}},
  \bibinfo{author}{\bibfnamefont{C.~R.} \bibnamefont{Simovski}},
  \bibnamefont{and} \bibinfo{author}{\bibfnamefont{S.~A.}
  \bibnamefont{Tretyakov}}, \\
\bibinfo{journal}{Phys. Rev. B}
  \textbf{\bibinfo{volume}{67}}, \bibinfo{pages}{113103}
  (\bibinfo{year}{2003}),\\
  \XDOI{10.1103/PhysRevB.67.113103}.

\bibitem[{\citenamefont{Gratus and McCormack}(2015)}]{Gratus-M-2015jo}
\bibinfo{author}{\bibfnamefont{J.}~\bibnamefont{Gratus}},
  \bibinfo{author}{\bibfnamefont{M.}~\bibnamefont{McCormack}},\\
  \bibinfo{journal}{J. Opt.} \textbf{\bibinfo{volume}{17}},
  \bibinfo{pages}{025105} (\bibinfo{year}{2015}),\\
  \XDOI{10.1088/2040-8978/17/2/025105}.

\end{thebibliography}


\end{document}